\documentclass{aa}
\usepackage{lineno}

\usepackage{graphicx}
\usepackage{txfonts}
\usepackage{amsmath}
\usepackage{xcolor}
\usepackage{soul}
\usepackage{upgreek}
\usepackage{comment}
\usepackage{url}
\usepackage{hyperref}
\usepackage{orcidlink}

\graphicspath{{./}{}}

\begin{document}

   \title{Dark and Luminous Matter in the Coma Cluster: Probing Galaxy Cluster Assembly Through Filaments with Weak Lensing and Multiwavelength Observations}

   \titlerunning{Dark and Luminous Matter in Coma}
   \authorrunning{HyeongHan et al.}

   \author{K. HyeongHan\inst{\ref{in:Duke},\ref{in:Yonsei}}\,\orcidlink{0000-0002-2550-5545}
           \and K. Finner\inst{\ref{in:IPAC}}\,\orcidlink{0000-0002-4462-0709}
           \and M. James Jee\inst{\ref{in:Yonsei},\ref{in:Davis}}\corrauth{mkjee@yonsei.ac.kr}\,\orcidlink{0000-0002-2550-5545}
           \and W. Lee\inst{\ref{in:Yonsei}}\,\orcidlink{0000-0002-2550-5545}
           \and Y. Jim\'{e}nez-Teja\inst{\ref{in:IAA},\ref{in:ON}}\,\orcidlink{0000-0002-6090-2853}
           \and S. Cha\inst{\ref{in:Yonsei}}\,\orcidlink{0000-0001-7148-6915}
           \and W. Kang\inst{\ref{in:Chicago}}\,\orcidlink{0009-0006-8665-5754}
           \and H. S. Hwang\inst{\ref{in:SNU},\ref{in:SNU2},\ref{in:SNU3}}\,\orcidlink{0000-0003-3428-7612}
           \and H. Cho\inst{\ref{in:Yonsei},\ref{in:CGER}}\,\orcidlink{0000-0001-5966-5072}
           \and E. Churazov\inst{\ref{in:MPA},\ref{in:IKI}}\,\orcidlink{0000-0002-0322-884X} 
           \and I. Khabibullin \inst{\ref{in:RPC},\ref{in:MPA},\ref{in:IKI}}
           \and N. Lyskova\inst{\ref{in:IKI},\ref{in:ASC}}\,\orcidlink{0000-0003-4917-7803}
           \and R. Sunyaev\inst{\ref{in:IKI},\ref{in:MPA}}\,\orcidlink{0000-0002-2764-7192}
           \and A. M. Bykov\inst{\ref{in:Ioffe}}
           }

   \institute{Department of Physics, Duke University, Durham, NC 27708, USA \label{in:Duke}
         \and Department of Astronomy, Yonsei University, 50 Yonsei-ro, Seoul 03722, Korea \label{in:Yonsei}
         \and IPAC, California Institute of Technology, 1200 E California Blvd., Pasadena, CA 91125, USA \label{in:IPAC}
         \and Department of Physics, University of California, Davis, One Shields Avenue, Davis, CA 95616, USA \label{in:Davis}
         \and Instituto de Astrof\'{i}sica de Andaluc\'{i}a–CSIC, Glorieta de la Astronom\'{i}a s/n, E–18008 Granada, Spain \label{in:IAA}
         \and Observat\'{o}rio Nacional, Rua General Jos\'{e} Cristino, 77 - Bairro Imperial de S\~{a}o Crist\'{o}v\~{a}o, Rio de Janeiro, 20921-400, Brazil \label{in:ON}
         \and Department of Astronomy \& Astrophysics, University of Chicago, 5640 S Ellis Avenue, Chicago, IL 60637, USA \label{in:Chicago}
         \and Astronomy Program, Department of Physics and Astronomy, Seoul National University, 1 Gwanak-ro, Gwanak-gu, Seoul 08826, Republic of Korea \label{in:SNU}
         \and SNU Astronomy Research Center, Seoul National University, 1 Gwanak-ro, Gwanak-gu, Seoul 08826, Republic of Korea \label{in:SNU2}
         \and Institute for Data Innovation in Science, Seoul National University, Seoul 08826, Republic of Korea \label{in:SNU3}
         \and Center for Galaxy Evolution Research, Yonsei University, 50 Yonsei-ro, Seoul 03722, Korea \label{in:CGER}
         \and Max Planck Institute for Astrophysics, Karl-Schwarzschild-Str. 1, D-85741 Garching, Germany \label{in:MPA}
         \and Space Research Institute (IKI), Profsoyuznaya 84/32, Moscow 117997, Russia \label{in:IKI}
         \and Rudolf Peierls Centre for Theoretical Physics, Department of Physics, University of Oxford, Clarendon Laboratory, Parks Rd, Oxford, OX1 3PU, United Kingdom \label{in:RPC}
         \and AstroSpace Centre of P.N. Lebedev Physical Institute of the Russian Academy of Sciences, Profsoyuznaya 84/32, Moscow, 117997, Russia \label{in:ASC}
         \and Ioffe Institute, Politekhnicheskaya st. 26, Saint Petersburg 194021, Russia \label{in:Ioffe}
              }

   \date{Received XXX; accepted XXX}

   \abstract
   % context heading 
   {The Coma cluster (Abell 1656; $z=0.023$) is a nearby rich galaxy cluster and a key laboratory for studying cluster assembly in the Cosmic Web.}
   % aims heading 
   {We characterize the projected dark matter distribution of Coma and its connection to galaxies, the intracluster medium, and reported intracluster filaments.}
   % methods heading 
   {We present a weak-lensing (WL) analysis of wide-field ($\sim$12-deg$^2$) Subaru/Hyper Suprime-Cam imaging. 
   We reconstruct the two-dimensional mass distribution, fit Navarro-Frenk-White (NFW) models, derive an aperture mass densitometry profile, and compare the WL signal with optical spectroscopy, eROSITA X-ray observations, radio data, and gas mass fraction diagnostics.}
   % results heading
   {A single-halo NFW fit yields $M_{200\mathrm{c}} = 8.2 \pm 0.7\times10^{14}\,M_{\odot}$. 
   The aperture mass profile agrees with the best-fit NFW model and the X-ray hydrostatic mass at $R\gtrsim20'$ ($\sim$560~kpc), suggesting that the current merger configuration does not strongly bias the global WL mass estimate.
   In contrast, the inner region shows a substantial hydrostatic bias ($b\lesssim0.5$). 
   A two-halo NFW fit centered on NGC~4874 and NGC~4839 gives $M_{200\mathrm{c}}=7.8\pm0.6\times10^{14}\,M_{\odot}$ and $0.9\pm0.2\times10^{14}\,M_{\odot}$, respectively, implying a $\sim$1:8 minor merger. 
   Using the gas mass fraction as a proxy for the merger state, we predict that the system is returning from first apocenter.
   We find a positive spatial correlation between the WL signal and the X-ray surface brightness, strongest along the reported intracluster filament (ICF) directions ($110^{\circ}$ and $340^{\circ}$), where shear-selected subhalos are predominantly detected.
   Finally, the $r$-band mass-to-light ratio ($M/L_r$) of Coma is radially constant with $\langle M/L_r\rangle \simeq 250 \pm 66 ~ M_{\odot} / L_{\odot}$ within $R_{200c}$, whereas the northern and western ICFs show substantially higher values of $M/L_r\sim1,000 ~ M_{\odot} / L_{\odot}$, suggesting that the ICFs are more strongly dark matter dominated than the cluster.}
   % conclusions heading 
   {Our results indicate that Coma is an active node of the cosmic web and demonstrate that joint WL and multiwavelength analyses can effectively probe cluster assembly and the dark matter content of ICFs.}

   \keywords{galaxies: clusters: individual: Abell 1656 -- galaxies: clusters: intracluster medium -- gravitational lensing: weak -- large-scale structure of Universe}

   \maketitle

\section{Introduction}

In the standard $\Lambda$ cold dark matter ($\Lambda$CDM) cosmology, structure grows hierarchically through anisotropic gravitational collapse driven by the primordial tidal field \citep{Zeldovich1970, Peebles1980, Bond1996Natur}. 
This process gives rise to the {\it Cosmic Web}, a network of voids, sheets, filaments, and halos that is ubiquitously seen in cosmological simulations \citep[e.g.,][]{Klypin1983, Colberg2005, Springel2006Natur, Bond2010} and mapped by large galaxy redshift surveys \citep[e.g.,][]{Huchra1982, deLapparent1986, Colless2001, York2000, Driver2009}. 
Among these components, filaments are particularly important because they contain a substantial fraction of the matter in the universe while occupying only a small fraction of the cosmic volume \citep[e.g.,][]{Hahn2007, Cautun2014, Libeskind2018}.
At the intersections of filaments, galaxy clusters form and continue to grow by accreting matter preferentially along the large-scale structure (LSS) \citep[e.g.,][]{vanHaarlem1993, Knebe2004, Rost2021, Kuchner2020}. 
Therefore, a holistic view of clusters and their surrounding filaments is essential for understanding cluster assembly, environmental processing of galaxies, and the coevolution of dark and baryonic matter \citep[e.g.,][]{Shandarin1984, Shandarin1989, Pimbblet2004, Colberg2005, Kartaltepe2008, Gonzalez2010, Noh2011, Kuchner2020}.

The Coma cluster, one of the nearest \citep[$z=0.023$;][]{Struble1999} rich galaxy clusters \citep{Abell1989}, is an active hub for such a study \citep[e.g.,][]{Williams1981, Fontanelli1984, Mahajan2010, Gavazzi2011, Seth2020, Einasto2025}. 
It has been investigated extensively across the electromagnetic spectrum and is known to host a variety of substructures. 
Optical photometric and spectroscopic studies have revealed complex dynamics, including substructures and color variations in the intracluster light (ICL), which are likely remnants of ongoing or recent mergers among member galaxies \citep[e.g.,][]{Adami2005_ICL, Jimenez-Teja2019, Jimenez-Teja2025}, as well as dynamical substructures inferred from line-of-sight velocities \citep[e.g.,][]{Colless1996, Adami2005, Gerhard2007, Adami2009, Healy2021, Jimenez-Teja2025, Kang2025}, and associations with the surrounding LSS \citep[e.g.,][]{Falco2014, Mahajan2018, Malavasi2020}.
On the other hand, X-ray observations have uncovered shocks, contact discontinuities, turbulence, and stripped gas associated with the infall of the NGC~4839 group \citep[e.g.,][]{Neumann2001, Neumann2003, Sanders2013, Simionescu2013, Zhuravleva2019NatAs, Churazov2021, Churazov2023}. 
Radio studies have further demonstrated the presence of diffuse synchrotron emissions, including a giant radio halo and radio relic generated by the merger activity \citep[e.g.,][]{Brown2011, Bonafede2021, Bonafede2022, Healy2021}. 
These multiwavelength results collectively show that Coma is not a relaxed monolithic halo, but a dynamically active system embedded in a rich large-scale environment \citep[e.g.,][]{Gregory1978, Fontanelli1984, Kim1989Natur, Gavazzi1999, Mahajan2010, Mahajan2018, Malavasi2020}. 
Yet, despite this wealth of observations, the projected dark matter distribution of Coma and its connection to the luminous structures remain incompletely understood.

Weak-lensing (WL) analyses of nearby ($z\ll0.1$) clusters such as Coma are intrinsically difficult because the lensing efficiency is low. 
However, this disadvantage is partly overcome by several important factors. 
Because the cluster subtends a very large solid angle, one can obtain a large number of background galaxies per unit physical area, which improves the statistical precision of the shear measurement \citep[e.g.,][]{Jee2017, HyeongHan2024NatAs}. 
In addition, the background galaxies can be selected with high purity \citep[$\gtrsim99\%$;][]{HyeongHan2025NatAs, Finner2025}, and the noise of unrelated LSS projected along the line of sight is reduced compared to studies of more distant clusters \citep[e.g.,][]{Hoekstra2001, Hoekstra2003, Hoekstra2011, Wu2019}. 
These advantages make Coma a compelling target for a high-fidelity WL analysis, provided that observational WL systematic uncertainties are well controlled.

There have been previous WL studies of the Coma cluster, but they have all been limited by either shallow imaging depth (i.e., low source density) or insufficient radial coverage in the cluster outskirts. 
Using Sloan Digital Sky Survey (SDSS) Data Release 5 \citep[DR5;][]{Adelman-McCarthy2007} imaging, \citet{Kubo2007} measured the cluster mass over a wide field ($R\sim14$~Mpc) and obtained $M_{200c}=1.88^{+0.65}_{-0.56}\times10^{15}~h^{-1}~M_{\odot}$\footnote{We indicate $R_{\Delta}$ as the radius where the average density becomes $\Delta$ times the critical density of the universe. Accordingly, $M_{\Delta}$ is the total mass within $R_{\Delta}$.}, but the low source density ($\sim1~\rm arcmin^{-2}$) led to large statistical uncertainty and likely enhanced the susceptibility to projection effects. 
With deeper CFHT/MegaCam imaging data, \citet{Gavazzi2009} achieved a source density of $\sim$10 arcmin$^{-2}$ and derived $M_{200c}=5.1^{+4.3}_{-2.1}\times10^{14}~h^{-1}_{70}~M_{\odot}$, but their analysis was limited to half of the cluster's $R_{200c}$ ($0.5R_{200c}$$\sim$1~Mpc).
Subaru-based studies using Suprime-Cam improved the source density further ($\sim$20--40 arcmin$^{-2}$) and revealed dark matter substructure associated with central galaxies and the infalling NGC~4839 group \citep{Okabe2010, Okabe2014}. 
In particular, the wider and deeper analysis of \citet{Okabe2014} investigated the properties (i.e., mass, truncation radii, and mass-to-light ratios) of the dark matter subhalos as a function of clustocentric distance; these subhalos were examined in detail by \cite{Kang2025} focusing on the correlations between weak-lensing and dynamical properties of the subhalos.
However, no prior WL analysis has simultaneously achieved a source density of $\sim$38 arcmin$^{-2}$ together with coverage beyond the virial radius \citep[$R_{vir}$;][]{Bryan1998, Colossus2018}, a regime in which the cluster’s connection to the surrounding LSS remains largely unexplored.

As a step toward addressing this, \citet{HyeongHan2024NatAs} used the matched-filter technique \citep{Maturi2013} to detect intracluster filaments (ICFs) through a WL analysis of Subaru/Hyper Suprime-Cam \citep[HSC;][]{HSC1, HSC2, HSC3, HSC4} data.
They reported the $3\sigma$ detection of ICFs along the northern (N; 110$^\circ$) and western (W; 340$^\circ$) directions whose mass overdensities are $\rho/\rho_{bkg} \sim 300$, where $\rho_{bkg}$ is the mean matter density, and southeastern (SE; 240$^\circ$) direction as a candidate.
Based on these detections, they argued for a connection between the ICFs and the surrounding large-scale ($\sim$10~Mpc) filaments identified by galaxy spectroscopy \citep[e.g.,][]{Mahajan2018, Malavasi2020}, suggesting that the ICFs are channels for matter inflow from the LSS.
This study qualitatively linked the WL-detected ICFs to baryonic tracers of the surrounding large-scale structure; however, a quantitative comparison between the dark and baryonic components has not yet been performed.

In this paper, we compare the dark matter component inferred from our WL analysis with the baryonic component in the Coma cluster traced by multiwavelength observations to better understand its assembly through the filamentary structures.
The WL analysis is based on archival Subaru/HSC imaging covering $\sim$12 deg$^2$, extending beyond the virial radius.
This wide-field dataset enables both an $M_{200c}$ mass measurement without extrapolating the shear profile and a two-dimensional reconstruction of the projected mass distribution over the cluster's $R_{vir}$.
In combination with the multiwavelength data, this dataset allows us to investigate the following:
i) the merger phase of the accreting NGC~4839 group \citep[e.g.,][]{Burns1994, Colless1996, Neumann2001, Neumann2003, Lyskova2019, Churazov2021}, 
ii) how the projected dark matter distribution correlates with the X-ray and galaxy-density structures along ICFs, and
iii) the dark matter and baryonic content of the reported ICFs.
We combine archival galaxy spectroscopy and X-ray datasets with our WL analysis to quantitatively address the connection between the projected dark matter distribution, the baryonic structures, and the ICFs.

This paper is organized as follows. 
In Section~\ref{sec:obs}, we describe the observations and the construction of the source catalog. 
Section~\ref{sec:wl} summarizes the WL formalism and mass reconstruction method. 
In Section~\ref{sec:mass_map}, we present the two-dimensional mass distribution and discuss the identification of the cluster center. 
We then compare the WL results with the optical, X-ray, and radio properties of Coma and its substructures in Section~\ref{sec:luminous}. 
Section~\ref{sec:mass} presents the mass estimation using parametric and non-parametric approaches, and subsequent subsections discuss the hydrostatic bias, gas fraction, and the infalling NGC~4839 group. 
In Section~\ref{sec:ml}, we discuss the mass-to-light ratios of the cluster and ICFs. 
We summarize our conclusions in Section~\ref{sec:conclusions}.

We adopt cosmological parameters of $H_0=70~\rm km~s^{-1}~Mpc^{-1}$, $\Omega_{m}=0.3$, $\Omega_{\Lambda}=0.7$ under a flat $\Lambda$CDM cosmology throughout this paper.
At the redshift of the Coma cluster ($z=0.023$), the plate scale is $0.466~\rm kpc~arcsec^{-1}$.

\section{Observations and Source Catalog} \label{sec:obs}

The data reduction, point-spread function (PSF) modeling, galaxy shape measurement, and source selection processes are presented in \cite{HyeongHan2024NatAs}. 
Here, we briefly introduce the dataset that we utilized, and refer readers to \cite{HyeongHan2024NatAs} for more details.
The observations were performed in 2016 and 2017 in HSC-{\it g} and -{\it r} filters, with average seeing values of 1\farcs3 and 1\farcs4, respectively.
Since the shape measurement is susceptible to the PSF dilution \citep{Bernstein2002}, we created a stacked image that is optimized for galaxy shape measurements by only retaining frames with $\left< \rm FWHM \right> < 0\farcs85$.
In addition, we excluded CCDs containing defective channels (\texttt{DET-ID} = 0, 9, 33, and 43) and those with more than 50\% vignetting.
The resulting final mosaic image has an average seeing of 0\farcs7, as measured from the stars.

\begin{figure}[!t]
    \centering
    \includegraphics[width=1\columnwidth]{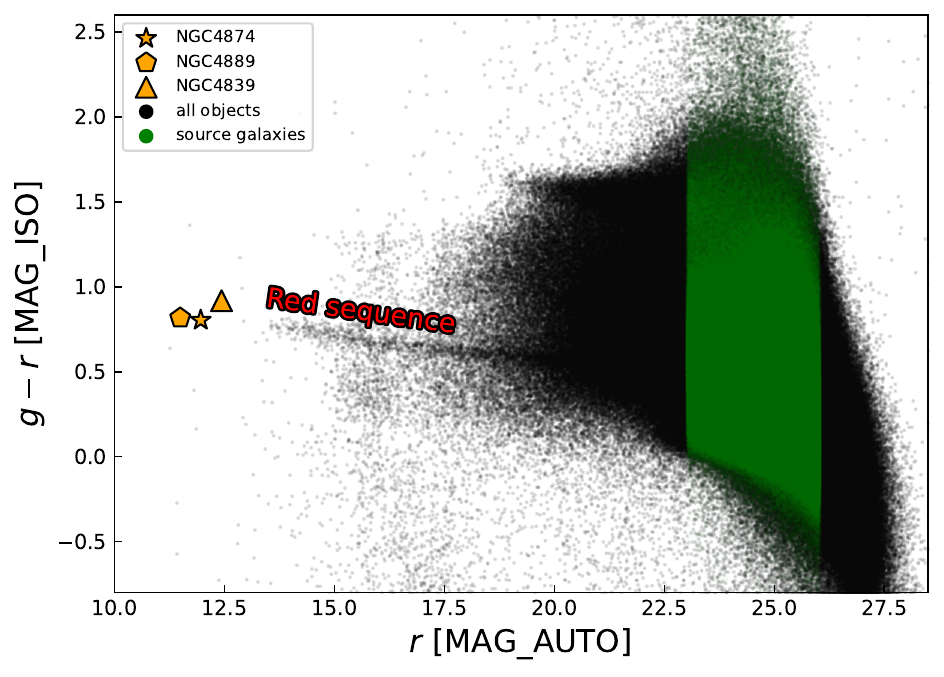}
    \caption{Color--magnitude diagram of the Coma cluster.
             Notable member galaxies NGC~4874, NGC~4889, and NGC~4839 are indicated by gold star, pentagon, and triangle, respectively.
             Green points mark the source galaxies used in the WL analysis in this study.
             Black points show detected objects in the field, with sources having \texttt{FLUX\_RADIUS} $< 3$ pix excluded for clarity.
             The red sequence is clearly visible, supported by the large number of spectroscopically identified member galaxies \citep[e.g.;][]{Jimenez-Teja2025, Kang2025}.
            }
    \label{fig:cmd}
\end{figure}

We measured galaxy shapes with the forward-modeling approach using the PSF-convolved elliptical Gaussian model \citep[e.g.,][]{HyeongHan2020, HyeongHan2024, HyeongHan2024NatAs, HyeongHan2025NatAs}. 
To model PSFs, we performed PCA analysis frame by frame using $\sim$47 stars per chip on average.
We interpolated the principal components using a third-order polynomial function and reconstructed the PSF model at the location of source galaxies.
Readers are referred to \cite{Jee2007} and \cite{Jee2011} for the implementation details.
As a diagnostic of the PSF smearing, we calculated the PSF ellipticity residual ($e^{star}-e^{model}$) and ($e^{resi}-e^{gal}$) cross-correlation functions.
These correlations have small amplitudes ($<10^{-6}$) on angular scales greater than $1'$ \citep{HyeongHan2024NatAs} which are negligible for this study.
After galaxy shape measurements, we selected source galaxies by imposing criteria on galaxies' magnitude ($23<r<26$), shape ($e<0.9$), size (semi-minor axis $b > 0.4$ pix, \texttt{FLUX\_RADIUS}$>0.5$ arcsec), and goodness of fit (\texttt{STATUS}=1, ellipticity measurement noise $\sigma_{mea}<0.4$).
The selected source galaxies are shown in the color--magnitude diagram (Figure~\ref{fig:cmd}).
The final number density of the source catalog is $\sim$38\,arcmin$^{-2}$.

\section{Weak-Lensing Method} \label{sec:wl}

\subsection{Basic Lensing Formalism}

Weak gravitational lensing refers to the tidal deflection of the light path induced by density inhomogeneities along the line of sight.
It causes the modification of the shapes of the background objects on the image plane \citep[for reviews, see][and references therein]{Schneider2005, Mandelbaum2018} described by the Jacobian transformation {\bf A}:
\begin{equation} 
    \bf{A} =  
    \begin{bmatrix}
    1 - \kappa - \gamma_1 & -\gamma_2 \\
    -\gamma_2             & 1 - \kappa + \gamma_1 
    \end{bmatrix},
    \label{eq:jacobianmatrix}
\end{equation} 
where $\gamma_{1(2)}$ is the complex shear, which causes anisotropic shape distortion, and $\kappa$ is the convergence, which represents the projected mass density in units of the critical surface density.
We often express the shear in complex notation $\gamma = \gamma_1 + i\gamma_2$.
The projected mass density $\Sigma$ can be found from $\Sigma=\kappa\Sigma_c$ by normalization with the lensing critical density, 
\begin{equation}
    \Sigma_c = \frac{c^2 D_s}{4\pi G D_l D_{ls}},
    \label{eq:crit_sigma}
\end{equation}
where $c$ is the speed of light, $G$ is the gravitational constant, $D_l$ is the angular diameter distance to the lens, $D_{ls}$ is the angular diameter distance from the lens to the source, and $D_s$ is the angular diameter distance to the source.
Since galaxy shapes are invariant under the isotropic $(1-\kappa)$ rescaling in the lensing Jacobian, WL shape measurements yield the reduced shear, $\text g=\gamma/(1-\kappa)$, rather than $\gamma$ alone.
In the WL regime ($\kappa \ll 1$), the approximation $\text g \approx \gamma$ holds.

To conveniently express both magnitude and direction of the shear, we define tangential ($\text g_+$) and cross ($\text g_{\times}$) shear components for a given reference point as
\begin{equation}
    \begin{aligned}
        \text g_+ = -\text g_\mathrm{1}  \rm cos (2 \it \phi) - \text g_\mathrm{2}  \rm sin (2 \it \phi) \\
        \text g_{\times} = \text g_\mathrm{1}  \rm sin (2 \it \phi) - \text g_\mathrm{2}  \rm cos (2 \it \phi) ,
    \end{aligned}
\end{equation}
where $\text g_1$ and $\text g_2$ are the measured ellipticity components of a source galaxy, and $\phi$ is the position angle of the major axis of the object.
Ideally, the averaged cross shear should be consistent with a null signal because of parity symmetry \citep{Schneider2003}.
Therefore, the excess in the cross shear component provides a diagnostic for the effect of residual systematics.

A cluster potential is not solely responsible for the observed shape distortion in a galaxy cluster field, but the integration of the cluster lensing signal, intrinsic galaxy shape, and the lensing signal caused by the LSS.
We obtain the reduced shear by weight averaging the raw ellipticity ($e^{raw}_{1,2}$) over an annulus $\theta$ as follows:
\begin{equation}
    \text g_{1,2}(\theta) = m\frac{\sum e^{raw}_{1,2} w_i}{\sum w_i},
\end{equation}
where we adopt the global multiplicative calibration factor $m=1.15$ \citep{HyeongHan2024NatAs}.
The inverse variance weight $w_i$ is given by
\begin{equation}
    w_i = \frac{1}{\sigma_{int}^2+\sigma^2_{mea, i}},
\end{equation}
where $\sigma_{int}$ is the intrinsic shape noise for which we adopt $\sigma_{int}=0.25$. 
The tangential shear uncertainty caused by the uncorrelated LSS $\sigma_{LSS}$ \citep{Hoekstra2001} is of order $10^{-3}$ at the clustocentric distance $10'< R < 100'$, which is negligibly small compared to the imposed shape noise.
In the case of the correlated LSS, we further discuss its effect in \textsection\ref{subsec:corrLSS}.

\begin{figure*}[!ht]
    \centering
    \includegraphics[width=\textwidth]{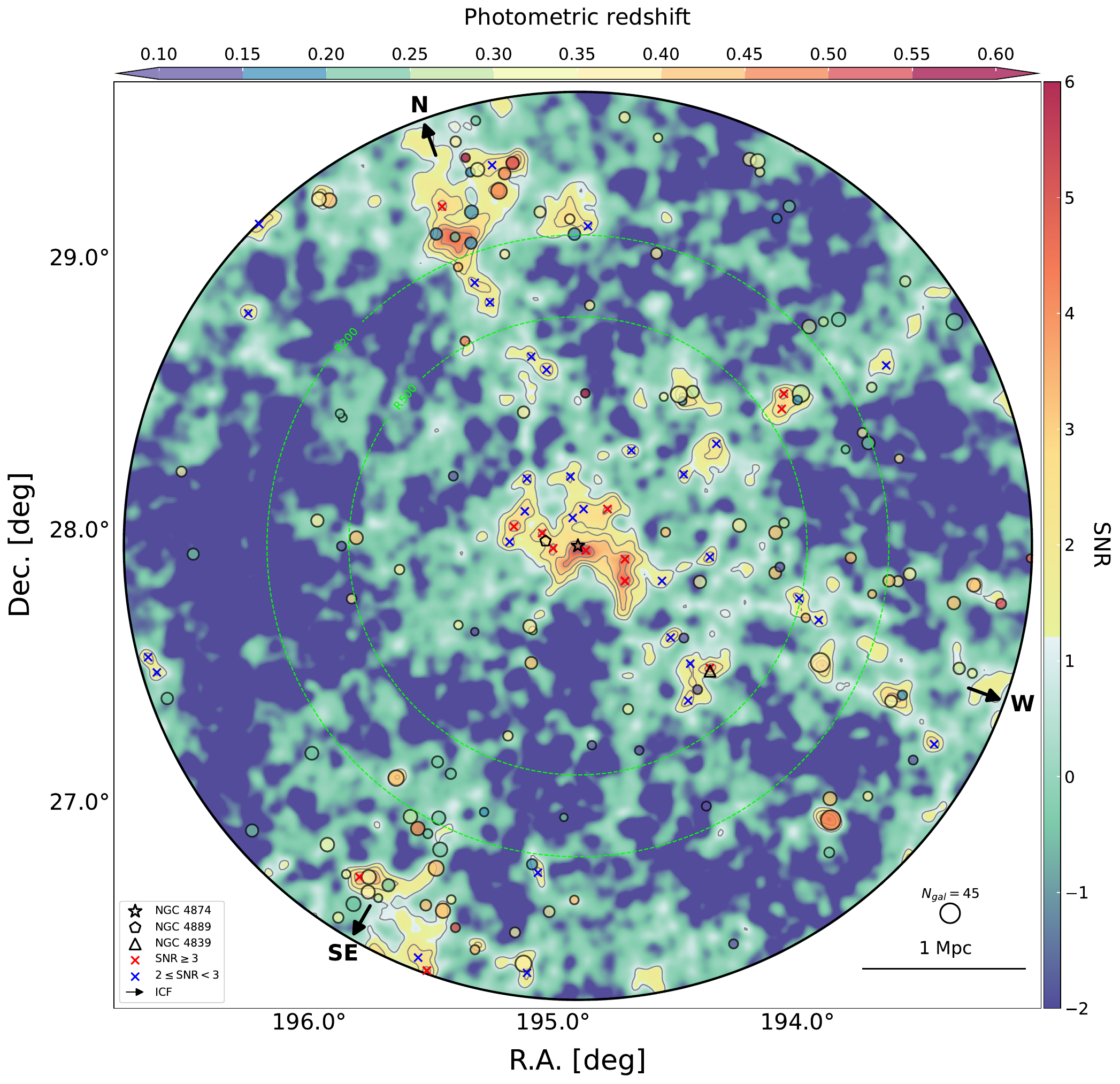}
    \caption{Two-dimensional mass distribution of the Coma cluster within a 2.8~Mpc-radius aperture centered on NGC~4874. 
    The background is the signal-to-noise ratio (SNR) map estimated using 1,000 realizations of the convergence ($\kappa$) map obtained by bootstrapping the source catalog.
    The gray contour starts from the $1\sigma$ level with $1\sigma$ step each. 
    The open star, pentagon, and triangle shapes indicate NGC~4874, NGC~4889, and NGC~4839, respectively. 
    The black arrows indicate the directions of ICFs reported in \cite{HyeongHan2024NatAs}.
    The inner and outer green dashed circles indicate $R_{500c}$ and $R_{200c}$, respectively. 
    The galaxy groups in the background are marked by filled circles with sizes linearly proportional to the number of members, and the photometric redshift is color coded. 
    The size of the open circle in the lower right corner corresponds to a background group with $N_{gal}=45$, which has the largest number of galaxies among the identified background groups.
    The red and blue crosses mark shear-selected subhalos with $\mathrm{SNR} \geq 3$ ($N=13$) and $2 < \mathrm{SNR} < 3$ ($N=44$), respectively.
    }
    \label{fig:mass_map}
\end{figure*}

\subsection{Projected Mass Mapping Technique}

A traditional approach to reconstruct the convergence field from a shear catalog is the direct Fourier inversion method (\citealt{Kaiser1993}; hereafter KS93). 
However, it yields a biased result due to the following well-known issues.
First, KS93 provides the conversion between the shear and convergence, while the observable is the reduced shear \citep[e.g.,][]{Schneider1995, Bartelmann2001}. 
Applying the transformation to the observed shear catalog, by definition, inherits a bias.
Second, the assumption $\text g \approx \gamma$ holds in the weak limit ($\kappa \ll 1$), however, it is no longer valid near deep gravitational potential wells \citep[e.g., at the central region of massive clusters;][]{Schneider1995, Seitz1997, Bartelmann2001}. 
Even if the assumption is valid, the approximation causes measurable biases at small angular scales that require a non-linear transformation \citep[e.g.,][]{White2005, Deshpande2020}.
Third, it applies a smoothing kernel to suppress noise in the observed shear, producing a smoothed convergence map \citep{Bartelmann2001}.
Fourth, KS93 is a non-local transformation that is subject to the limited observation window and masked regions \citep[e.g.,][]{Schneider1995, Seitz1996, Jeffrey2018}.
The blinded shear information introduces artifacts and/or biases near such boundaries.
Lastly, the ambiguity in the transformation [$\kappa \rightarrow \lambda \kappa + (1 - \lambda)$], so-called ``mass-sheet degeneracy'', makes the shear-convergence conversion a non-unique problem \citep[e.g.,][]{Falco1985, Schneider1995, Bartelmann2001, Bradac2004}.

To mitigate the aforementioned limitations, we reconstruct the convergence map using a deep-learning method proposed by \cite{Hong2021} and improved by \cite{Cha2025} for wide-field WL mass reconstruction. 
This method presented by \citet{Cha2025} outperforms KS93 in terms of noise suppression, centroid recovery, dynamical range, and mass estimation. 
Here, we briefly introduce this deep-learning technique and refer readers to \cite{Cha2025} for more details.

\cite{Cha2025} trained a convolutional neural network (CNN) using 42,000 data-augmented convergence maps generated from the MassiveNuS cosmological simulation\footnote{https://skiesanduniverses.org/Simulations/Guest/MassiveNu/} \citep{Liu2018}.
The network takes three $512\times512$ pix$^2$ input channels: two mean ellipticity components ($e_1$ and $e_2$) and the ellipticity measurement noise ($\sigma_{mea}$) which are tailored to match the shear measurement of the Coma cluster using the Subaru/HSC data in \cite{HyeongHan2024NatAs}.
Then, the CNN predicts the convergence maps covering $3.5\times3.5$~deg$^2$ on a $512\times512$ pixel grid ($\sim$$0\farcm41$ per pixel). 
The CNN architecture preserves the full resolution of the map and is designed to recover both compact high $\kappa$ peaks and diffuse LSS.

To assess the performance of the reconstruction, they presented several diagnostics using 200 test data sets. 
For instance, a pixel-by-pixel comparison between the true and reconstructed convergence maps reveals a strong correlation, with a total least-squares slope of $\sim$$0.7$. 
In particular, the reconstructed $\kappa$ distribution agrees well with the true distribution over the range $-0.1 < \kappa < 0.1$. 
In addition, the comparison of the $\kappa$--$\kappa$ autocorrelation functions demonstrates that the CNN reconstruction successfully recovers structures on both small ($\geq0\farcm82$) and large scales. 
A discrepancy appears at scales smaller than 2 pixels ($\sim$$0\farcm8$), which is attributed to the finite source density and the resulting resolution limit of the reconstruction. 
Therefore, we smooth the CNN-predicted mass map with a Gaussian kernel whose standard deviation ($\sigma$) corresponds to 24\farcs6, or 11.5~kpc.

\section{Two-dimensional Mass Distribution} \label{sec:mass_map}

WL measures the line-of-sight projected matter density contrast, so the reconstructed mass map is inherently a projected surface-density field weighted by the lensing kernel.
It contains the target cluster signal plus contributions from uncorrelated LSS (i.e., LSS noise) and any additional foreground or background overdensities along the same sightline \citep[e.g.,][]{Hwang2014}. 
Since the Coma cluster is nearby and spans a few degrees on the sky, we must be careful not to mistake the contribution of line-of-sight weak-lensing signal as that from the cluster. 
We therefore use the extensive multiwavelength data available for Coma to identify and separate these interloping structures.

\subsection{Shear-peak detection} \label{subsec:shear_peak}

To distinguish line-of-sight structures unrelated to Coma, we examine galaxy-group catalogs from the literature (NSC -- \citealt{Gal2003}; MaxBCG -- \citealt{Koester2007}; SDSSCGB -- \citealt{McConnachie2009}; WHL -- \citealt{Wen2009}; GMBCG -- \citealt{Hao2010}; MCXC -- \citealt{Piffaretti2011, Sadibekova2024}) and retain only groups with photometric redshifts $z_{\rm phot} < 0.01$ or $z_{\rm phot} > 0.04$, thereby excluding the redshift range that brackets Coma ($z=0.023$).
In some cases, galaxy groups identified in the literature show slight differences in their reported positions and membership assignments. 
To create uniformity, we merge them into a single entity if they are within a projected separation less than 4 arcmin and a redshift difference less than 500~km~s$^{-1}$.

Figure~\ref{fig:mass_map} shows our signal-to-noise ratio (SNR) map of the reconstructed convergence field with the previously reported background structures from the literature and identified shear peaks in this analysis. 
We use the {\tt photutils} package \citep{photutils} to detect shear peaks within $R<2.8$~Mpc, adopting a peak-detection box size of $2\farcm9$, comparable to the uncertainty in the mass centroid (see \textsection\ref{subsec:center}).
After removing the shear peaks that spatially overlap ($<2$ arcmin) with the structures from literature, we find that 57 shear peaks remain (peak SNR$>2$) with 13 of them having a peak SNR $\geq3$.
We note that the contribution from most of those background structures to the mass map is expected to be insignificant, owing to their low lensing efficiency and modest average richness ($\langle N_{\rm gal}\rangle\sim15$). 
Therefore, our shear-peak catalog provides a conservative selection.

The remaining global shear-peak (or subhalo) distribution shows radially elongated features, particularly toward the northern (N) and western (W) ICFs \citep[][]{HyeongHan2024NatAs}.
This is consistent with the picture that Coma is embedded in the large-scale filamentary environment of the Coma supercluster, in which the W filament connects toward the Leo (Abell~1367) cluster \citep[e.g.,][]{Williams1981, Fontanelli1984, Kim1989Natur, Gavazzi1999, Cybulski2014, Mahajan2010, Mahajan2018, Malavasi2020}, and the N-extension traces the continuous LSS reported in the galaxy spectroscopic studies \citep{Mahajan2018, Malavasi2020}.
In addition, the northwest (NW; 30$^{\circ}$) candidate ICF\footnote{We classify it as a candidate owing to its relatively low significance of $\sim$2.5$\sigma$ in \cite{HyeongHan2024NatAs}.} exhibits a spatial alignment of subhalos, consistent with the presence of a less prominent filamentary structure.
On the other hand, the number of subhalo candidates is substantially lower on the eastern side of the main halo, with the southeastern (SE) direction dominated by background structures.

\subsection{Where is the Center of the Coma cluster?} \label{subsec:center}

\begin{figure}[!t]
    \centering
    \includegraphics[width=1\columnwidth]{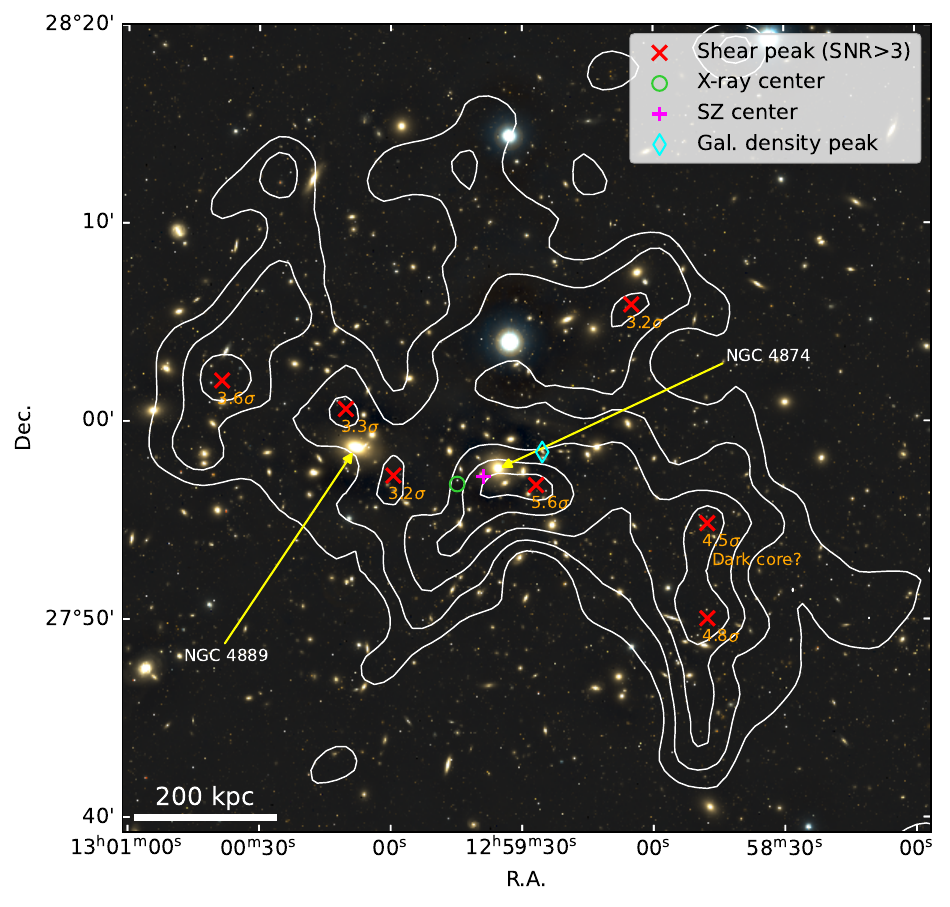}
    \caption{Central region of the Coma cluster. 
             The background HSC pseudo-color composite image is overlaid with white WL SNR contours starting from $1\sigma$ with $1\sigma$ step each. 
             Red crosses denote shear peaks with $\mathrm{SNR}>3.5$ detected in \textsection\ref{subsec:shear_peak}, and the orange numbers give their peak significances. 
             The centroid uncertainty of the convergence peak associated with NGC~4874 is roughly $3'$ (84~kpc).
             The galaxy density peak, X-ray center, and SZ center are marked by the cyan diamond, green circle, and magenta plus, respectively. 
             The locations of NGC 4874 and NGC 4889 are labeled. 
             The scale bar corresponds to 200 kpc.
             }
    \label{fig:center}
\end{figure}

In a merging cluster, the definition of the cluster center is often non-trivial, because the peaks traced by the X-ray emission, the thermal Sunyaev-Zel'dovich effect \citep[tSZ;][]{Sunyaev1972}, and galaxy number density frequently do not coincide. 
The brightest cluster galaxy (BCG) is commonly adopted as a proxy for the potential minimum \citep{Beers1983}, even in merging systems \citep{Finner2025}, however, the presence of two comparably bright, closely spaced central galaxies, NGC~4874 and NGC~4889, makes the choice of the cluster center in Coma less straightforward.
NGC~4889 is identified as the BCG by definition, has a radial velocity consistent with that of the surrounding cluster population \citep{Gerhard2007, Sanders2020}, and exhibits the highest intracluster globular cluster surface density \citep{Madrid2018}.
However, both the X-ray centroid \citep[RA, Dec = 194.9367, 27.9472$^{\circ}$;][]{Sato2011} and the centroid of the Compton-$y$ parameter measured from the Planck tSZ signal \citep[RA, Dec = 194.9118, 27.9537$^{\circ}$;][]{PlanckCollaborationXXVII2016} are measurably offset from NGC~4889 by 151 and 185~kpc, respectively\footnote{We note that the uncertainties of the X-ray and tSZ centroids are as large as $\sim$10$'$ (280~kpc).}. 
Meanwhile, NGC~4874 is $\sim$0.4 mag fainter than NGC~4889 in the $r$ band, yet lies closer to the X-ray and tSZ centroids (by 59 and 22 kpc, respectively), hosts a radio source \citep[e.g.,][]{Willson1970, Lal2022, Murgia2024}, and is associated with a larger amount of ICL than around NGC~4889 \citep{Jimenez-Teja2025}.
These properties are characteristic of a central galaxy at the bottom of the potential well, underscoring the longstanding ambiguity in defining the dynamical center of the Coma cluster.

As the two BCGs complicate the definition of the Coma cluster center (Figure~\ref{fig:center}), we explore the convergence field to examine the gravitational potential.
The strongest ($5.6\sigma$) convergence peak lies 53~kpc east of NGC~4874.
To estimate the centroid uncertainty of the mass peak, we analyze convergence peak distributions using the 1,000 realizations of bootstrapped convergence maps \citep[e.g.,][]{Finner2017, Finner2023, HyeongHan2024}.
We then construct a centroid probability density map by smoothing the bootstrapped peak-centroid distribution with a Gaussian kernel of $\sigma=24\farcs6$, corresponding to the angular resolution of the CNN training set.
This procedure is repeated using top-hat priors with widths ranging from 60 to 1,000~kpc.
We persistently find that NGC~4874 is consistent with a $5.6\sigma$ convergence peak within $3\sigma$ centroid uncertainties ($\sim$3$'=84$~kpc). 
In the case of NGC~4889, it is associated with $3.3\sigma$ and $3.2\sigma$ convergence peaks that are consistent within the $2\sigma$ centroid uncertainties.
Both peaks are offset from NGC~4889 by $\sim$50~kpc which may reflect the past collision between the two cD galaxies \citep[e.g.,][]{Colless1996, Adami2005, Gerhard2007, Andrade-Santos2013, Gu2020, Jimenez-Teja2025}.
Given the spatial coincidence with the strongest convergence peak and the proximity to the X-ray and SZ peaks, we set the location of NGC~4874 as the cluster center throughout this paper.

\section{Comparison with luminous matter} \label{sec:luminous}

\begin{figure*}[!th]
    \centering
    \includegraphics[width=\textwidth]{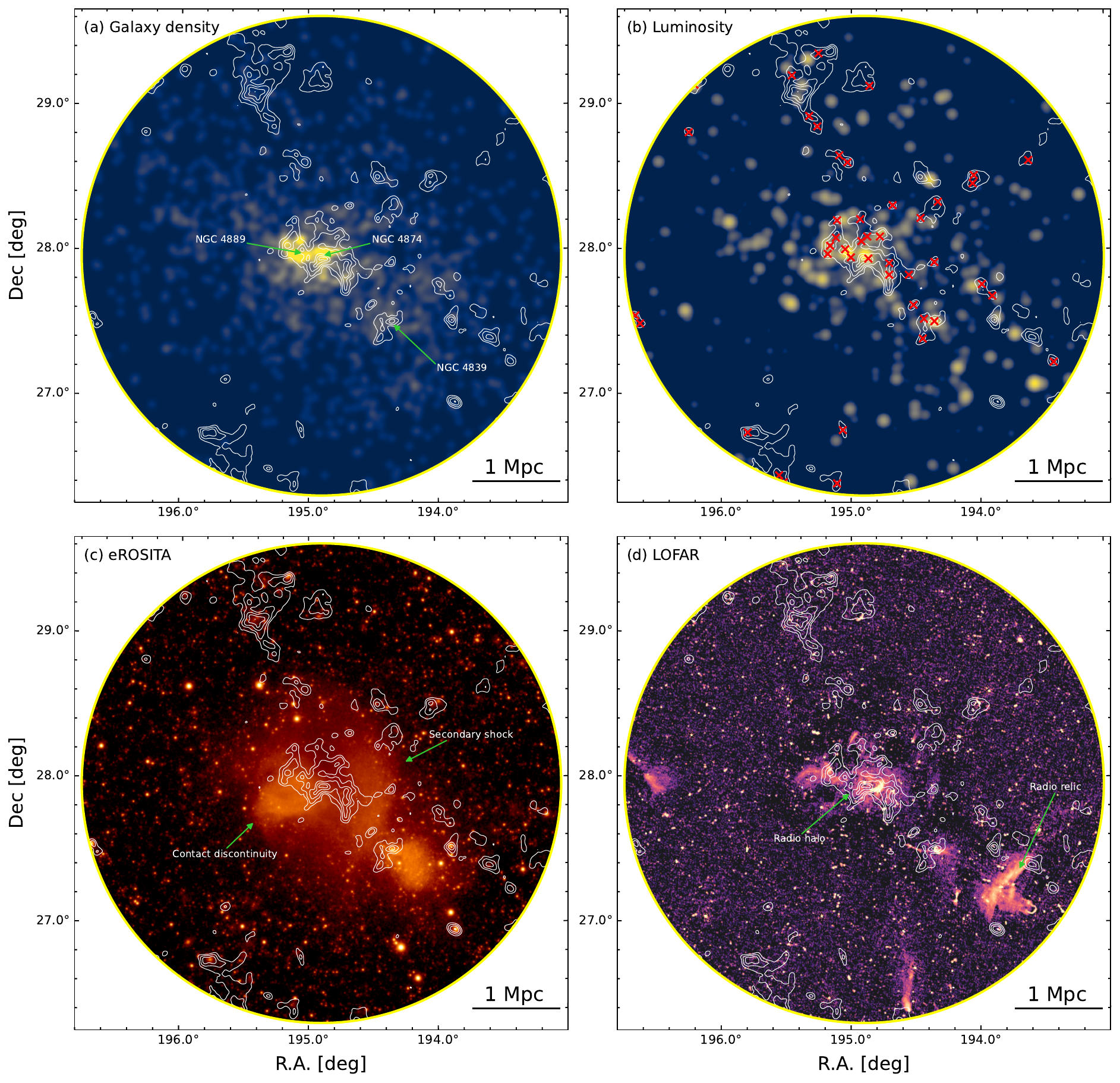}
    \caption{Mass distribution overlaid on multiwavelength observations within a radius of 2.8~Mpc centered at NGC~4874.
             The white contours represent the signal-to-noise ratio of the reconstructed mass that start from $1\sigma$ level with $1\sigma$ step each. 
             Top left: The background is smoothed ($\mbox{FWHM}$$\sim$$0\farcm6$) galaxy density map using spectroscopically confirmed member galaxies \citep{Kang2025} on a linear scale. 
             It shows that galaxy overdensity is extended toward the direction of Abell~1367 where the western ICF is detected \citep{HyeongHan2024NatAs}. 
             We note that the spectroscopic completeness near the south-eastern field is below 50\%. 
             The central brightest galaxies and NGC~4839 are marked with green arrows. 
             Top right: Smoothed galaxy luminosity map on a logarithmic scale. 
             The red crosses denote the $>2\sigma$ shear peaks that are detected in the projected mass map.
             Bottom left: Flat-fielded eROSITA X-ray surface brightness map \citep{Churazov2021} in log scale. 
             It features a contact discontinuity and a secondary shock whose locations are marked by the green arrows. 
             The first passage of the NGC~4839 group is believed to be responsible for generating the secondary shock and contact discontinuity \citep{Lyskova2019, Zhang2021, Churazov2021, Churazov2023}, while its ram-pressure stripped tail indicates ongoing infall.
             The shear peak ($3.9\sigma$) is located at NGC~4839, leading the ram-pressure stripped tail of its group.
             Bottom right: The publicly available LOFAR DR2 continuum imaging \citep{Shimwell2022} in log scale.
             The green arrows indicate the locations of radio halo and relic \citep[e.g.,][]{Brown2011, Bonafede2022}.
             }
    \label{fig:multi}
\end{figure*}

\begin{figure*}[!t]
    \centering
    \includegraphics[width=\textwidth]{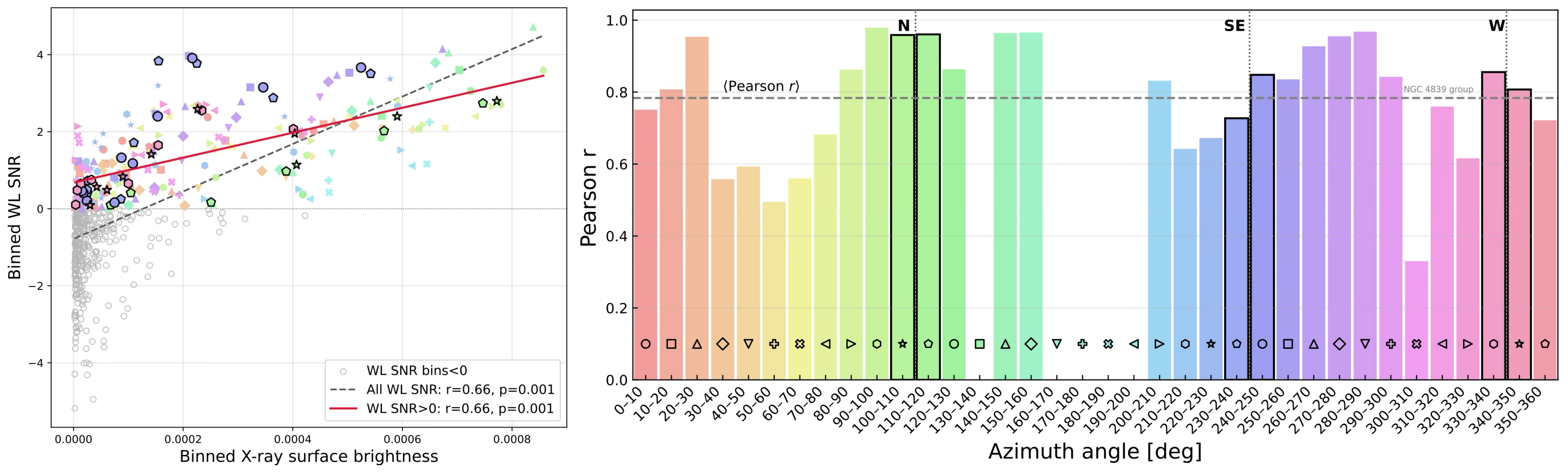}
    \caption{Comparison between the binned X-ray surface brightness (SB) and the binned weak-lensing (WL) signal, and their dependence on azimuthal direction. 
             Left: Binned WL signal-to-noise ratio (SNR) as a function of binned X-ray surface brightness. 
             Gray open circles denote bins with WL SNR $< 0$, while the colored symbols represent measurements from individual azimuthal sectors. 
             The dashed gray line shows the best-fit relation obtained using all bins, whereas the solid red line shows the fit restricted to bins with WL SNR $> 0$.
             The corresponding Pearson correlation coefficients, $r$, and $p$-values are listed in the legend and suggest the mild correlation between the X-ray SB and WL signals.
             Each colored marker corresponds to different azimuthal bin shown in the right panel.
             Bins corresponding to the ICF directions marked in the right panel are highlighted with black edges.
             Right: Pearson correlation coefficient computed in azimuthal sectors with an opening angle of $10^\circ$.
             Coefficients estimated using more than three bins are shown.
             The horizontal dashed line marks the mean of the correlation coefficients. 
             Black-outlined bars highlight sectors that are aligned with the direction of N--, SE--, and W--ICFs, which are showing relatively strong correlations.
             }
    \label{fig:wl_xray_corr}
\end{figure*}

Figure~\ref{fig:multi}(a) and (b) show the member galaxy density and luminosity distributions of the Coma cluster overlaid with the mass contours, respectively.
On large scales, the galaxy density and luminosity maps are spatially correlated with the mass map.
The mass distribution exhibits elongation, particularly toward the north and west, consistent with matter accretion along the ICFs.
We find no clear signature of a galaxy overdensity along the SE--ICF that was previously reported by \cite{HyeongHan2024NatAs} as a candidate.
However, the spectroscopic completeness near the southeastern field is below 50\%, and therefore the galaxy distribution in this region should be interpreted with caution.
A chain of galaxies has also been reported to the east of the Coma center based on spectroscopic surveys \citep{Mahajan2018, Malavasi2020}, however, we do not detect a corresponding dark matter structure in this direction.

Figure~\ref{fig:multi}(c) and (d) present the mass map contours over the flat-fielded\footnote{The X-ray surface brightness map is divided by a sum of the simple beta-model and a flat background as it is described in \cite{Churazov2021}.} X-ray surface brightness map \citep{Churazov2021} obtained by deep observations with eROSITA telescope \citep{Predehl2021} onboard the SRG observatory \citep{Sunyaev2021} and the LOFAR DR2 continuum image \citep{Shimwell2022}.
The X-ray and radio emissions trace the thermal and non-thermal components of the ICM and highlight clear features of past merging activity. 
For example, the past collision between the central subhalo and the NGC~4839 group \citep[e.g.,][]{Lyskova2019, Zhang2021, Churazov2021, Churazov2023} has produced a contact discontinuity and a secondary shock, both of which are now detached from the cluster core. 
In addition, turbulence in the ICM has formed a radio halo that qualitatively is in spatial agreement with the mass overdensity in the central halo \citep{Bonafede2022}.
Another noticeable feature is the NGC~4839 group, infalling toward the central halo of Coma, which is currently undergoing a merger generating the radio relic \citep[e.g.,][]{Burns1994, Neumann2001, Brown2011, Lyskova2019, Mirakhor2020, Bonafede2021, Bonafede2022, Churazov2021}. 
We further discuss its merger phase in \textsection\ref{subsec:ngc4839} utilizing the WL and X-ray data.

\subsection{WL--X-ray correlation}

The X-ray observations of hot gas have been useful tracers of galaxy clusters \citep[e.g.,][]{Borgani2001Natur, Rosati2002}.
Numerical simulations \citep[e.g.,][]{Yoo2024, Yoo2025} further suggest that the dark matter and X-ray emission within galaxy clusters exhibit a strong two-dimensional spatial correspondence. 
However, observational comparisons between these components have so far been limited to coarse spatial binning, owing to either the low source density of WL data or insufficient X-ray photon counts \citep[e.g.,][]{Eckert2015Natur, Jauzac2016, Finner2021, Finner2025}.

We compare the eROSITA X-ray surface brightness (SB) map \citep{Churazov2021, Churazov2023} with our two-dimensional mass map of the Coma cluster to investigate how closely the ICM traces the projected mass distribution.
To quantify the spatial correspondence of the diffuse X-ray emission, we use the background-subtracted, point-source masked, Gaussian ($\sigma=30''$) smoothed X-ray image \citep{Churazov2021}.
We then bin both the X-ray SB and WL SNR maps using the same set of 3 arcmin ($\sim$80~kpc) radial annuli out to 50 arcmin ($\sim$1.4~Mpc) in azimuthal sectors with an opening angle of 10$^{\circ}$ and compute the mean X-ray SB and mean WL SNR in each two-dimensional bin. 
We restrict the analysis to radii within $50'$ to avoid the noise-dominated outer region of the X-ray SB map and the WL signals from the background structures.

We compare the binned X-ray SB and WL signal in the left panel of Figure~\ref{fig:wl_xray_corr}.
It shows a strong positive correlation regardless of whether negative WL signals are included.
When all bins are included, we obtain a Pearson correlation coefficient of $r=0.66$ with a $p$-value of $p=0.001$, indicating a statistically significant correspondence between the projected mass and the ICM. 
Since the measurements are correlated across bins, we estimate the $p$-value empirically from the 1,000 bootstrap realizations.
Restricting the analysis to bins with positive WL SNR yields a comparable correlation ($r=0.66$, $p=0.001$). 
This suggests that regions of enhanced X-ray emission generally coincide with regions of stronger projected mass.

We further investigate the directional dependence of the dark matter--X-ray correlation to probe the connection to the LSS.
The right panel of Figure~\ref{fig:wl_xray_corr} shows that the WL--X-ray correlation varies depending on azimuthal direction.
The strongest correlations are found in sectors at azimuthal angles of 100$^{\circ}$-120$^{\circ}$, 140$^{\circ}$-160$^{\circ}$, 270$^{\circ}$-290$^{\circ}$, and 330$^{\circ}$-350$^{\circ}$.
They are aligned approximately with the N-- (110$^{\circ}$), SE-- (240$^{\circ}$), and W-- (340$^{\circ}$) ICFs \citep{HyeongHan2024NatAs}, where the Pearson coefficient is above the azimuthal average. 
The correlation over a wide range of azimuthal angles along the SE--ICF is in agreement with the presence of a broad structure as previously noted by \cite{HyeongHan2024NatAs}.
However, this interpretation should be treated with caution because the structure has an unusually large width and shows an apparent absence of member galaxies along this direction (Figure~\ref{fig:multi}), although the spectroscopic completeness in the corresponding region is below 50\% \citep{Kang2025}.
The 140$^{\circ}$-160$^{\circ}$ sector was not reported as one of the ICFs, yet a galaxy overdensity is reported in large-scale studies \citep{Mahajan2018, Malavasi2020} and \cite{Cha2025} detected a filament candidate from the convergence map using {\tt DisPerSE} \citep{Sousbie2011a, Sousbie2011b}.
In addition, the candidate ICF at $30^{\circ}$ also shows a strong correlation.
Overall, this correspondence can be interpreted as mass accretion occurring along the preferred direction where the filaments are located.

\subsection{The presence of a dark core?}

One would expect galaxies to be found in dark matter halos and the shear-selected subhalos show good agreement with the density and luminosity maps in Figure~\ref{fig:multi}.
One notable exception is the shear peak located $\sim$13$'$ west of NGC~4874 (RA, Dec = 194.6940, 27.9101$^{\circ}$) showing an extension toward the south (Figure~\ref{fig:center}). 
It shows a high ($4.5\sigma$) peak significance that is comparable to the one associated with NGC~4874, however, it is devoid of a comparably bright central galaxy and lacks significant X-ray excess emission detected from the eROSITA observation \citep[e.g.,][]{Churazov2021, Churazov2023}.
This significant shear peak is also found in the previous WL mass maps presented by \cite{Gavazzi2009}, \cite{Okabe2010}, and \cite{Okabe2014}. 
With the absence of a bright central galaxy, the peak is reminiscent of a ``dark core'' reported by \cite{Jee2012, Jee2014} in A520 or dark structures \citep{Kim2026} from the Dark Energy Survey Y3 data \citep{Sevilla-Noarbe2021}.

\cite{Adami2009} explained its presence as a background massive group in the line of sight, putatively as part of the SDSS Great Wall \citep{Gott2005}.
With deep spectroscopy using VIMOS, they found an unvirialized population of faint galaxies ($r>21$ mag; $N=58$) at $z=0.054$ overlapping with the G8 and G9 Coma substructures defined in \citet{Adami2005} on the projected plane.
From the absence of the X-ray excess emission, they placed an upper limit of the substructures on the mass to be less than $5\times10^{12}~M_{\odot}$.
Although \citet{Jimenez-Teja2025} reported stellar emission from the ICL and identified a galaxy group at the dark-core position based on the DESI EDR spectroscopy catalog \citep{DESI2022, DESI2024} using the DS+ algorithm \citep{Dressler1988, Benavides2023}, the system has only four member galaxies and lacks a central bright galaxy whose magnitude is expected to be comparable with NGC~4874 and/or NGC~4889. 
This is at odds with the peak's high WL significance.

Meanwhile, the existence of a high-z cluster can be excluded by the spectroscopic and photometric studies \citep[e.g.,][]{Adami2005, Adami2009}.
In addition, the background massive group ($z=0.054$) reported by \cite{Adami2009} spatially coincides with the putative filament area reported in \cite{Malavasi2020}. 
However, they do not draw a solid conclusion because the two studies probe different physical scales when identifying filamentary structure along the line of sight.
Therefore, these pieces of evidence do not fully explain the presence of the dark core and the question remains unresolved.

\section{Mass Estimation of the Coma cluster} \label{sec:mass}

The mass of the Coma cluster has been weighed by various techniques, including X-ray \citep{Hughes1989}, SZ \citep{PlanckCollaborationXXVII2016}, galaxy velocity dispersion \citep[e.g.,][]{Rines2003, Sohn2017, Ho2022NatAs, Jimenez-Teja2025}, and WL \citep{Kubo2007, Gavazzi2009, Okabe2010, Okabe2014}.
Although WL serves as a direct probe of cluster masses, limited depth and/or incomplete sky coverage of the previous studies resulted in inconsistent mass measurements ($M_{200c}\sim4\text{--}19 \times 10^{14}~h^{-1}~M_{\odot}$).
In this section, we estimate the mass of the Coma cluster with parametric and non-parametric approaches using the wide-field imaging that covers the cluster's virial radius.

\subsection{Parametric approach} \label{subsec:nfw}

\begin{figure}[!t] 
    \centering
    \includegraphics[width=1\columnwidth]{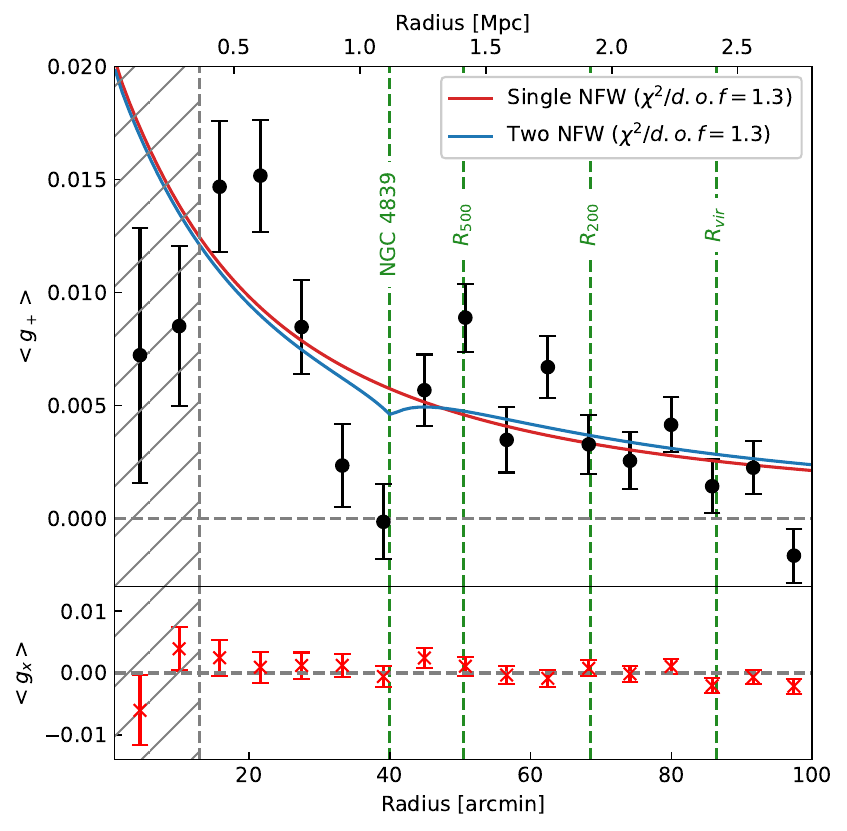}
    \caption{Reduced shear profiles of the Coma cluster. 
             Top: Tangential shear ($g_{+}$) radial profile. 
             The red solid line is the best-fit single NFW profile ($\chi^2/d.o.f=1.3$) assuming the $c$--$M$ relation \citep{Ishiyama2021} and the blue solid line indicates the best-fit result of the two-NFW-halo fitting ($\chi^2/d.o.f=1.3$) which is motivated by the presence of the NGC~4839 group. 
             The green lines indicate the locations of NGC~4839, $R_{500c}, R_{200c}$ and $R_{vir}$. The hatched region is excluded during the analysis. 
             The error bars indicate a $1\sigma$ uncertainty that accounts for the shape, measurement, and the large-scale structure noises. Bottom: Cross shear ($g_{\times}$) radial profile. It is consistent with the null signal.}
    \label{fig:tangential_shear}
\end{figure}

Figure~\ref{fig:tangential_shear} shows the azimuthally averaged tangential ($g_+$) and cross shear ($g_\times$) profiles with the best-fit Navarro-Frenk-White profile \citep[NFW;][]{NFW1996, NFW1997} models.
The $g_{\times}$ component, which diagnoses the remaining residual systematic error, is consistent with null.
We estimate the cluster mass by fitting an NFW profile \citep{Wright2000} to our unbinned tangential shear measurement through $\chi^2$ minimization.
We center the profile at NGC~4874 as it spatially coincides with the center of the cluster potential (\textsection~\ref{subsec:center}), and assume the $c$--$M$ relation \citep{Ishiyama2021} to break the degeneracy between the scale radius and concentration of the NFW profile.
We discard the central $\sim$360~kpc (or $13'$) region, where the value is motivated by \cite{Okabe2014}, to avoid bias due to the mis-centering, contamination of the BCG and member galaxies, non-linear shear response, and small-scale baryonic effects.
The best-fit NFW model yields a Coma cluster mass of $M_{200c} = 8.2\pm0.7\times10^{14}~M_{\odot}$ with a reduced chi-square of 1.3.
This result shows good agreement with the previous mass estimation by \cite{Okabe2014}. 
A corresponding concentration of $c_{200c}=3.82\pm0.01$ is derived from the adopted $c$--$M$ relation. 
We note that the quoted uncertainty on the concentration hereafter is obtained by propagating only the statistical uncertainty in the mass through the adopted $c$--$M$ relation. 
It therefore does not include the intrinsic scatter of the relation or other systematic uncertainties, and should not be interpreted as a direct measurement of the Coma concentration with this level of precision.

Motivated by the presence of the merging NGC~4839 group, we perform a two-halo NFW fit assuming that Coma is a superposition of two NFW halos.
The best-fit masses (concentrations) obtained using the $c$–$M$ relation, with the two halos centered on NGC~4874 and NGC~4839, are $M_{\mathrm{NGC4874},200c}=7.8\pm0.6\times10^{14}~M_{\odot}$ ($c_{\mathrm{NGC4874}, 200c}=3.83\pm0.01$) and $M_{\mathrm{NGC4839},200c}=0.9\pm0.2\times10^{14}~M_{\odot}$ ($c_{\mathrm{NGC4839},200c}=4.43^{+0.13}_{-0.09}$), respectively, with a reduced chi-square value of 1.3.
The result suggests that the cluster merger with the NGC~4839 group has a mass ratio of approximately 1:8.
Although the best-fit single- and two-halo NFW models yield similar reduced chi-square values, likely because the NGC~4839 component is much less massive than the main cluster, the two-halo model may provide a more physically appropriate description given the known presence of the NGC~4839 group.

\subsection{Non-parametric approach} \label{subsec:amd}

Adopting an analytic halo model inevitably introduces model bias into the inferred mass and concentration, because clusters do not necessarily follow the assumed density profile or $c$--$M$ relation. 
This issue becomes particularly important in merging galaxy clusters, where disturbed mass distributions and merger-driven structural evolution can significantly amplify the WL mass bias \citep{Lee2023, Euclid2024}.

To diagnose the potential halo model bias, we therefore employ the non-parametric aperture mass densitometry (AMD) method \citep[e.g.,][]{Fahlman1994, Kaiser1995b, Clowe2000, Jee2005}.
AMD computes the projected mass enclosed by a given aperture by comparing the density of an inner ($r<r_1$) region with a control annulus ($r_2<r<r_{max}$).
We define the control region as an annulus spanning radii from $r_2 = 80'$ (2.2~Mpc) to $r_{\max} = 100'$ (2.8~Mpc).
Since the input shear is the ``reduced'' shear, we update the convergence in the control annulus predicted by the single NFW profile ($\bar\kappa=0.001$) until it converges.
Finally, we estimate the non-parametric masses of Coma at 1 and 2~Mpc to be 
$4.3\pm 1.2 \times 10^{14}~M_{\odot}$ and $8.0\pm 4.9 \times 10^{14}~M_{\odot}$, respectively.
This shows good agreement with the parametric mass estimates enclosed within radii of 1 and 2~Mpc, $4.4\pm0.3\times 10^{14}~M_{\odot}$ and $7.6\pm0.5\times 10^{14}~M_{\odot}$, respectively.

\begin{figure}[!t] 
    \centering
    \includegraphics[width=1\columnwidth]{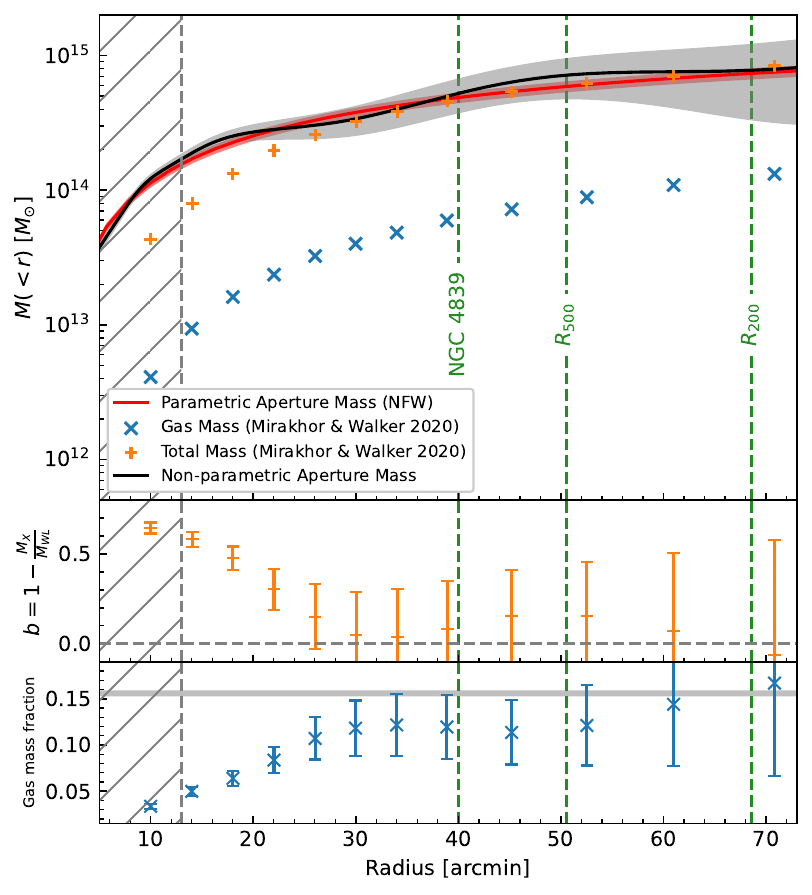}
    \caption{Projected mass profile of the Coma cluster. 
             Top: We compare the best-fit NFW mass (red) with the AMD profile (black). 
             Shaded regions are $1\sigma$ statistical uncertainties. 
             The hatched region is excluded when we estimate the parametric cluster mass.
             We adopted the X-ray derived gas (blue) and total (orange) mass from \cite{Mirakhor2020} marked in crosses. 
             There is no systematic difference between the best-fit NFW and non-parametric AMD mass profiles. 
             Middle: Hydrostatic bias ($b$) profile of the Coma cluster.
             It shows a significant bias at the central ($R\lesssim20'$) region while the hydrostatic equilibrium is reached at the outskirts.
             $M_X$ and $M_{WL}$ denote the X-ray derived total mass and the AMD WL mass, respectively.
             Bottom: Gas mass fraction of the Coma cluster. 
             The blue cross represents the gas mass fraction dividing the gas mass by the aperture mass. 
             The gray shaded region is the universal baryon fraction adopted from \cite{PlanckCollaborationXIII2016}. 
             The gas mass fraction at the cluster outskirts is consistent with the universal value.}
    \label{fig:amd}
\end{figure}

Figure~\ref{fig:amd} presents cumulative projected mass profiles from the AMD approach (black) in comparison with our best-fit single-NFW-halo model (red) under the assumption of the $c$--$M$ relation \citep{Ishiyama2021}.
One might have expected a noticeable discrepancy between the parametric and non-parametric mass estimates, given that Coma is undergoing a merger with the NGC~4839 group, 
However, the projected mass of the best-fit NFW halo shows excellent agreement with the AMD mass at all radii within the uncertainty.
This is because the WL mass bias induced by a merger is related to the merger configuration such as the mass ratio, initial velocity, impact parameter, and viewing angle \citep{Lee2023, Euclid2024}.
In the case of the Coma cluster, the low mass ratio of the merger ($\sim$0.1), the long time since collision \citep[$>1$~Gyr;][]{Lyskova2019}, and the large impact parameter \citep[e.g.,][]{Churazov2021} would have mitigated the merger-induced bias, allowing the agreement between the single-halo NFW model and the AMD mass profile.

\subsection{Effect of the correlated LSS} \label{subsec:corrLSS}

Galaxy cluster masses are a fundamental observable for cosmology, as the abundance of clusters and its evolution with redshift are sensitive to the growth of structure and the expansion history of the Universe \citep[e.g.,][]{Vikhlinin2009, Allen2011, Mantz2014, Bocquet2019, Costanzi2019, Abbott2020, Ghirardini2024, Lesci2025}.
However, clusters are not isolated systems; the correlated LSS contributes to the projected mass distribution in WL measurements and can therefore affect the inferred cluster mass \citep[e.g.,]{Becker2011, Bahe2012}.
The Coma cluster is no exception, as its outskirts are connected to several large-scale filaments \citep[e.g.,][]{Mahajan2018, Malavasi2020, HyeongHan2024NatAs} that may contribute non-negligibly to the measured shear signal.

To empirically assess the sensitivity of the cluster mass estimate to the filamentary structures, we repeat the WL mass estimation after masking the shear catalog in the directions of the identified ICFs.
The difference between the fiducial and masked measurements provides a measure of how strongly the inferred cluster mass depends on the filamentary sectors.
Because this procedure also removes part of the cluster shear signal in those directions, we interpret this comparison as a robustness test rather than a direct decomposition of the cluster and filament masses.
Specifically, we mask the shear field along the N--, SE--, and W--ICFs with opening angles of $30^{\circ}$, $40^{\circ}$, and $10^{\circ}$, respectively, and then perform single-halo NFW model fitting.
Depending on the adopted filament mask, the inferred $M_{200c}$ of the Coma cluster is up to $\sim$40\% lower than the fiducial mass.
As a complementary check, we also perform the opposite test by masking the non-filamentary regions and fitting the single-halo NFW model using only the shear field in the ICF sectors.
This fit yields an $M_{200c}$ that is $\sim$10\% higher than the fiducial mass.
The level of variation found in these masking tests is broadly consistent with theoretical expectations that correlated LSS can contribute significantly to the scatter in WL cluster mass estimates \citep{Becker2011}.
This result supports the interpretation that the shear signal along the ICF directions contributes positively to the fiducial cluster mass estimate, although it should not be interpreted as a direct measurement of the ICF mass.
We note that these tests suggest that the filamentary structures contribute non-negligibly to the WL-inferred cluster mass, but the exact contribution cannot be isolated by this masking approach alone.

The distinction between the cluster halo and its correlated filamentary structures is not always well defined \citep[e.g.,][]{Cen1997, Becker2011, Diemer2014}.
A detailed discussion of this ambiguity is beyond the scope of this paper and is deferred to future work.
In this paper, we therefore report the mass of the Coma cluster as presented in \textsection\ref{subsec:nfw} and \textsection\ref{subsec:amd}.

\subsection{Comparison with hydrostatic mass}

The hot ICM observed in X-rays has been widely used as a tracer of the gravitational potential of galaxy clusters \citep[e.g.,][]{Borgani2001Natur, Mantz2014} under the assumption of hydrostatic equilibrium.
However, the X-ray-derived mass can deviate from the WL mass because it is sensitive to the dynamical state of galaxy clusters \citep[e.g.,][]{Nagai2007, Piffaretti2008} and to additional non-hydrostatic effects, such as turbulence, bulk motions, and gas clumping \citep[e.g.,][]{Lau2009, Nagai2011, Simionescu2011Sci}.
To assess the impact of these effects, we compare Coma's mass profile inferred from X-ray observations and AMD.

Figure~\ref{fig:amd} presents the radial profiles of our WL mass together with the X-ray-derived gas mass (blue crosses) and total mass (orange pluses) from \citet{Mirakhor2020}, where the latter assumes hydrostatic equilibrium.
The comparison shows good agreement at larger radii ($R \gtrsim 20' \sim 0.6$~Mpc) for both the parametric and non-parametric reconstructions, although the AMD mass uncertainties remain large. 
Accordingly, the inferred hydrostatic bias is relatively small in the outer region, $b = 1 - M_{X}/M_{WL} \lesssim 0.1$, but becomes substantially larger in the inner region ($R \lesssim 20'$), reaching $b \lesssim 0.5$.
This large $b$ suggests that the ICM is not in hydrostatic equilibrium and that additional non-thermal support and/or merger-driven disequilibrium contributes to the pressure budget.
The non-thermal pressure in Coma is reported to be $\sim$3\%\footnote{We note that it does not include the full velocity field relative to the mean velocity of galaxies.} by recent XRISM Resolve analyses \citep{XRISMComa2025, Gatuzz2026}, while the coherent bulk motion is detected in the east-west direction.
Taken together, these results suggest that the low X-ray hydrostatic mass in Coma is driven primarily by large-scale merger-induced motions \citep[e.g.,][]{Neumann2003, Churazov2021} and the resulting departure from hydrostatic equilibrium.

The bottom panel of Figure~\ref{fig:amd} shows the gas fraction profile, which provides a useful diagnostic of the baryonic content of galaxy clusters.
The gas mass fraction increases with radius and approaches the cosmic baryon fraction, $f_{b}=0.156$ \citep{PlanckCollaborationXIII2016}, at the largest radii probed ($R_{200c}$).
Our result is consistent with previous observational and numerical studies reporting that the gas fraction rises with radius and tends toward the cosmic baryon fraction near the virial radius \citep[e.g.,][]{Gonzalez2013, Dvorkin2015, Angelinelli2023, Rasia2025, Matsushita2025arXiv}.

\subsection{The infalling NGC~4839 group} \label{subsec:ngc4839}

\begin{figure}[!t]
    \centering
    \includegraphics[width=1\columnwidth]{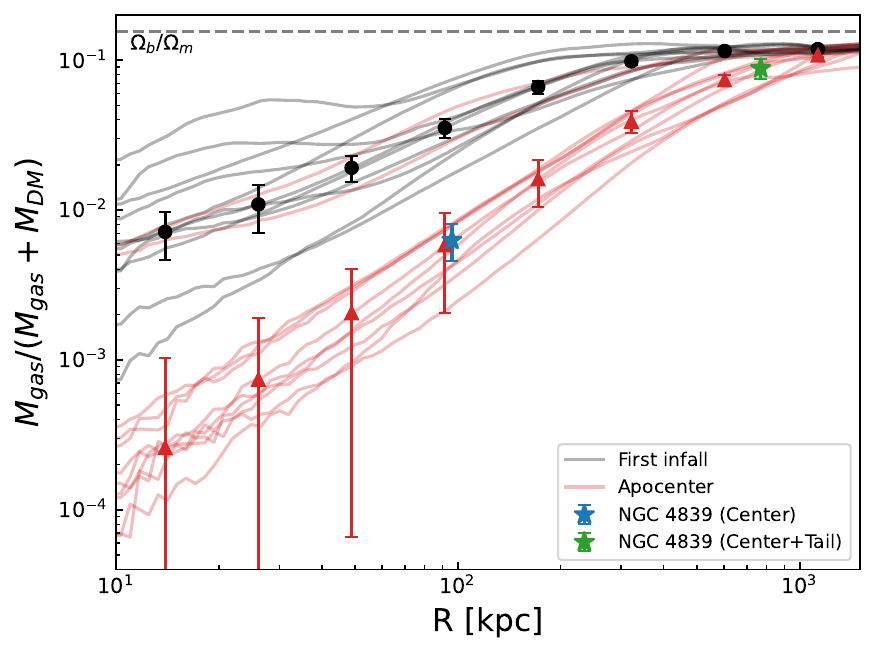}
    \caption{Gas mass fraction profiles of galaxy groups from the TNG-Cluster cosmological zoom-in simulation \citep[][]{2024A&A...686A.157N}. 
             The black solid lines indicate groups infalling for the first time at 1.1~Mpc separation. 
             The red solid lines indicate galaxy groups at the location of the first apocenter. 
             The circles and triangles represent the median value of the binned data with a $1\sigma$ statistical uncertainty. 
             The gray horizontal line refers to the universal baryon fraction adopted from \cite{PlanckCollaborationXIII2016}. 
             The stars represent the case of the NGC~4839 group using the gas \citep{Sasaki2016, Lyskova2019} and dark matter mass within a truncation radius ($r_t=3.43'\sim96$~kpc; blue) and 8 times the truncation radius (green), respectively. 
             The observed gas mass fraction of the NGC~4839 group suggests that the group has already passed its first core passage.
             }
    \label{fig:tngc}
\end{figure}

Despite the NGC~4839 group's relatively small mass, it is broadly shaping the hydrodynamic state of the Coma cluster, underscoring the importance of constraining its merger history.
The prominent stripped X-ray tail \citep[e.g.,][]{Neumann2001, Sasaki2016, Mirakhor2020, Mirakhor2023, Churazov2021, Churazov2023} strongly suggests that the group is infalling into the main halo; however, this alone does not uniquely determine its merger phase.
Earlier works based on the line-of-sight Coma galaxy motions \citep[e.g.,][]{Colless1996} and the X-ray distribution \citep[e.g.,][]{Neumann2001, Neumann2003} suggested that the NGC~4839 group is in its first infall through the western filament.
On the other hand, there are observational and theoretical studies suggesting that the NGC~4839 group is approaching its second core passage \citep[e.g.,][]{Burns1994, Biviano1996, Lyskova2019, Sheardown2019, Zhang2019, Churazov2021}. 
For example, the radio-halo front \citep{Bonafede2022}, the truncated atmosphere of the NGC~4839 group \citep{Lyskova2019, Sheardown2019}, and the presence of the bridges in X-ray, radio, ICL, and intracluster globular cluster analyses \citep[][]{Cho2016, Bonafede2021, Churazov2021, Oh2023, Jimenez-Teja2025} are all predicted to be produced during the NGC~4839 group's first core passage.
In addition, the radio relic \citep[e.g.,][]{Brown2011, Bonafede2022}, at a large projected separation ($\sim$2~Mpc) from the cluster center, is expected to be a runaway shock that is launched during its first infall \citep{Zhang2019}.

We use the gas mass fraction of the NGC~4839 group to constrain its merger phase, since the degree of stripping depends sensitively on the merger stage. 
Figure~\ref{fig:tngc} shows the gas mass fraction profiles of galaxy groups in the TNG-Cluster cosmological zoom-in simulation \citep{2024A&A...686A.157N} in comparison with the NGC~4839 group. 
From the simulation, we first selected the subhalos within a mass range of $[0.7,1.5]\times10^{14}\rm~M_{\odot}$ hosted by a central halo having a mass of $[7,15]\times10^{14}\rm~M_{\odot}$ from the merger catalog \citep{2024A&A...686A..55L}.
Then, we divided them into two categories depending on their merger phases, namely whether they are infalling to the main halo for the first time\footnote{We selected the first infallers at $\sim$1.1 Mpc separation in three-dimensional space which resembles the case of the NGC~4839 group.} (black; pre merger) or at their apocenter (red; post merger).
Each infall system shows a large variation as the effectiveness of the stripping significantly varies with its trajectory and gas density \citep[e.g.,][]{Roediger2007, Boselli2022}.
Nevertheless, the profiles demonstrate systematic differences at a given radius, reflecting the effects of gas stripping depending on the merger phase.

We estimate the gas mass fraction of the NGC~4839 group using the gas mass reported by \citet{Sasaki2016, Lyskova2019}.
At the truncation radius $r_t$ (i.e., $3.43'\sim96$~kpc), the group's gas mass fraction is in excellent agreement with the gas mass fractions measured near the apocenter.
This suggests that the NGC~4839 group is in a post-merger stage.
Given the post-merger signatures observed in previous studies \citep[e.g.,][]{Burns1994, Cho2016, Bonafede2021, Churazov2021, Oh2023, Jimenez-Teja2025}, this behavior supports the scenario in which the group is in its returning phase.
At large radii ($8r_t$; center+tail), the gas fraction is consistent with both scenarios and remains compatible with the universal baryon fraction \citep{PlanckCollaborationXIII2016}.
This demonstrates that the merger phase of the system can be constrained through joint analysis of WL and X-ray data.

\section{Mass-to-light ratio} \label{sec:ml}

\begin{figure}[!t]
    \centering
    \includegraphics[width=1\columnwidth]{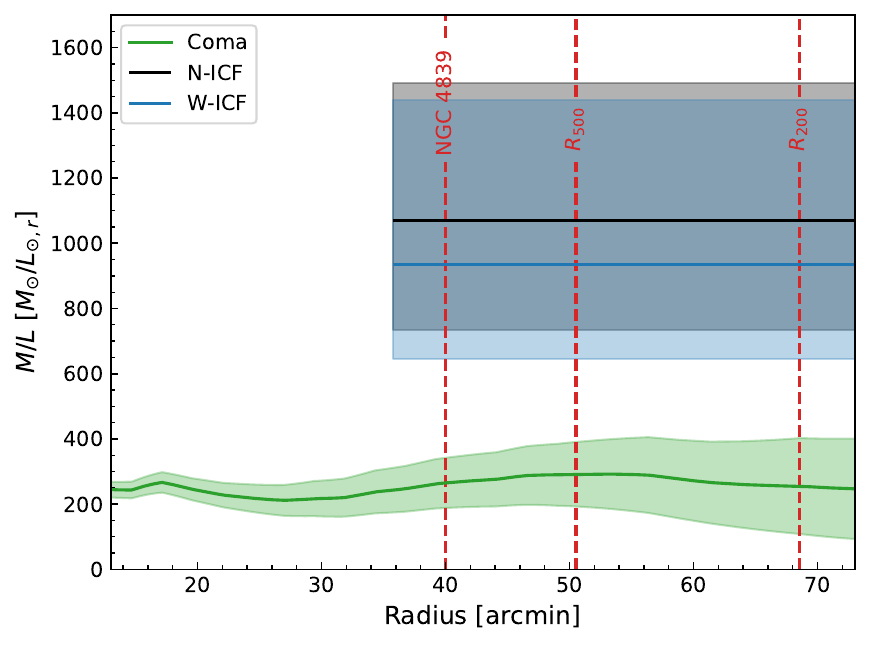}
    \caption{Projected $M/L$ profiles of the Coma cluster and the N-- and W--ICFs. 
             The mass is derived from the AMD and the luminosities are calculated using the member galaxies identified in \cite{Kang2025}.
             The shaded regions represent $1\sigma$ statistical uncertainties. 
             Red vertical dashed lines indicate the location of NGC~4839 and characteristic radii.
             The $M/L$ of ICFs are substantially higher than the host halo, suggesting that they are dominated by the dark matter component.
            }
    \label{fig:mass_to_light}
\end{figure}

Figure~\ref{fig:mass_to_light} shows the $r$-band cumulative mass-to-light ratio ($M/L_r$) profile of Coma, compared to those of the reported ICFs.
We adopt the projected mass estimated by the AMD profile (\textsection\ref{subsec:amd}) and calculate the SDSS $r$-band luminosity of member galaxies in units of the solar luminosity \citep{Willmer2018} from the Coma spectroscopy catalog \citep[i.e.,][]{Kang2025}.
The radial profile of the $M/L_r$ is flat up to $R_{200}$ with an average value of $250\pm66 ~ M_{\odot} / L_{\odot}$, which is consistent with previous reports \citep[e.g.,][]{Merritt1987, Okabe2014} for the Coma cluster. 
In contrast, other studies of major-merger galaxy clusters at $z\sim0.2$--$0.4$ find that $M/L$ initially increases with radius and then decreases at larger radii \citep{Kim2019, Finner2023}.

For the ICFs, we compute projected linear mass densities using the assumed profile, $\gamma_+ = \kappa_{2D} = \kappa_0 / [1 + (h/h_c)^2]$, where $\kappa_0$ is a normalization constant, and $h_c$ is a characteristic width.
We select a box-shaped area that is enclosed by $1 < R < 2.8$ Mpc with a width of $2h_c$, at the orientation reported in \cite{HyeongHan2024NatAs}.
The corresponding luminosity of the member galaxies is calculated within this same area.
The integrated linear mass densities in those regions are $1.91\times10^{14}~M_{\odot}~\rm Mpc^{-1}$, $0.78~\times10^{14}~M_{\odot}~\rm Mpc^{-1}$, and $3.40~\times10^{14}~M_{\odot}~\rm Mpc^{-1}$ for N--, W--, and SE--ICFs, respectively. 
Because the estimated ICF masses include both the filamentary component and the projected contribution from the host cluster, we subtract the projected mass of the best-fit NFW profile within the same aperture as a first-order correction to avoid overestimating the ICF masses.
As a result, we obtain $M/L_r$ values of $1069^{+423}_{-334} ~ M_{\odot} / L_{\odot}$, $934^{+504}_{-289} ~ M_{\odot} / L_{\odot}$, and $2164^{+1101}_{-634} ~ M_{\odot} / L_{\odot}$ for N--, W--, and SE--ICFs, respectively.
These ICF $M/L_r$ values are substantially higher than that of the cluster.
In particular, the SE--ICF exhibits the highest $M/L_r$ where we find a deficit of member galaxies along that direction; however, the spectroscopic completeness near the southeastern field is low ($<50\%$), so this deficit should be interpreted with caution.
In the following, we discuss the possible origin of these large $M/L_r$.

As the WL mass can be affected by projection effects, contribution from the host halo, and the assumed filament geometry, we compare it with an independent dynamical estimate.
The dynamical mass of the filaments can be calculated using velocity dispersions under the assumption of isothermality \citep{Eisenstein1997}.
Within the same filamentary regions defined above, we estimate the velocity dispersion of N--, W--, and SE--ICFs to be $736\pm83 ~ \rm km~s^{-1}$, $593\pm78 ~ \rm km~s^{-1}$, and $782\pm51 ~ \rm km~s^{-1}$, respectively. 
The corresponding $M/L_r$s are $880\pm200 ~ M_{\odot} / L_{\odot}$ (N), $1200\pm320 ~ M_{\odot} / L_{\odot}$ (W), and $1270\pm170 ~ M_{\odot} / L_{\odot}$ (SE) where N-- and W--ICFs are statistically consistent with the estimates from the weak lensing.
Although the dynamically derived SE--ICF's $M/L_r$ is a factor of two lower than the one estimated via WL, it is still significantly higher than the cluster value.

These high $M/L_r$ estimates suggest that the observed ICFs are dark matter dominated.
At face value, this appears to contradict theoretical expectations \citep[e.g.,][]{GaneshaiahVeena2019}, which predict that the dark-matter--to--stellar-mass ratio in filaments should be consistent with that in the cluster nodes given the same stellar population.
This discrepancy may be alleviated by the prediction that a large fraction of the halo population in filaments resides at low masses: roughly half of the total filament halo mass ($M_{\rm h}$) is contained in objects with $M_{\rm h} \lesssim 10^{11}~M_\odot$ \citep[e.g.,][]{Cautun2014, GaneshaiahVeena2021}, corresponding to stellar masses of $\lesssim 5\times10^{8}~M_\odot$ \citep[e.g.,][]{Zu2015, DiCintio2017, Sifon2018}. 
At the distance of the Coma cluster ($\sim$100~Mpc), such low-mass systems have $r$-band magnitudes of $\sim$20 for a single stellar population. 
Consequently, our luminosity measurement likely misses a substantial fraction of the stellar light associated with low-mass filament halos, leading us to underestimate the filament luminosity and therefore overestimate the observed $M/L_r$.
Testing this hypothesis will require deeper spectroscopic observations and an improved characterization of the star-formation histories of galaxies residing in the filaments.

Although ICFs are closer to the nodes which can exhibit different filament properties, there are a few reports on the $M/L$ of cosmic filaments.
For example, \cite{Schirmer2011} reported $M/L_i=305\pm201$\footnote{The subscript denotes the $i$-band luminosity.} by integrating the three main filaments in the supercluster SCL2243-0935 ($z=0.447$), one of the largest known systems at intermediate redshifts. 
On the other hand, \cite{Yang2020} derived $M/L_r=351\pm137$ by stacking luminous red galaxy pairs using the LOWZ ($0.15<z<0.43$) and CMASS ($0.43<z<0.7$) sample from the Baryon Oscillation Spectroscopic Survey \citep{Eisenstein2011, Dawson2013} and found no significant redshift evolution between the two bins.
Within the uncertainties, their filament $M/L$ measurements are broadly consistent with the universal values \citep{Loveday2015}.
Compared with these measurements of large-scale filaments, our $M/L$ estimate suggests that ICFs, which lie closer to the nodes of the cosmic web, may trace a structurally and physically distinct filament environment.

\section{Conclusions and Summary} \label{sec:conclusions}

In this paper, we presented a WL analysis of the Coma cluster based on archival Subaru/HSC optical imaging data.
The wide-field ($\sim$12-deg$^2$) coverage extends beyond the cluster virial radius, allowing us to reconstruct both the two-dimensional projected mass distribution and the azimuthally averaged mass profile.
By combining our WL results with optical spectroscopy, X-ray, and radio observations, we investigated how the dark matter distribution in Coma is related to its luminous components and ongoing assembly.
Our main results are summarized as follows:

\begin{itemize}
    \item We reconstructed the projected mass distribution of Coma using the CNN architecture.
    We identified 57 shear-selected subhalo candidates with WL SNR$>2$, whose spatial distribution extends toward the northern and western directions.
    This distribution is consistent with the previously reported ICFs that trace the surrounding LSS.
    
    \item The reconstructed convergence field supports NGC~4874 as the most plausible center of the Coma cluster.
    Although the central region is complicated by the presence of the two cD galaxies NGC~4874 and NGC~4889, NGC~4874 is more closely aligned with the dominant WL peak as well as the X-ray and tSZ centroids, the bright radio source, and the ICL distribution.

    \item We estimated the mass of the Coma cluster with parametric and non-parametric approaches.
    From the azimuthally averaged tangential shear profile, we obtained a best-fit single-halo NFW mass of
    $M_{200\mathrm{c}} = (8.2 \pm 0.7)\times10^{14}\,M_{\odot}$.
    A two-halo fit, motivated by the infalling NGC~4839 group, yields
    $M_{\mathrm{NGC4874},200\mathrm{c}}=7.8\pm0.6\times10^{14}\,M_{\odot}$ and $M_{\mathrm{NGC4839},200\mathrm{c}}=0.9\pm0.2\times10^{14}\,M_{\odot}$,
    implying a minor merger with a mass ratio of approximately 1:8.
    A comparison between the parametric NFW and the non-parametric AMD profiles shows excellent agreement over $R_{200c}$.

    \item The WL and X-ray projected mass profiles are broadly consistent at large radii.
    Beyond $R \gtrsim 20'$ ($\sim$0.6~Mpc), the hydrostatic and lensing masses agree well, indicating that the hydrostatic approximation is more valid in the outskirts. 
    It suggests that the hydrostatic mass is consistent with the lensing mass between $\sim 0.5 R_{500c}$ and $R_{200c}$, although the uncertainties are large, varying between 25 and 50\% across this range of radii. 
    By contrast, the central region exhibits a large and statistically significant hydrostatic bias, likely reflecting disturbed gas motions and departures from equilibrium caused by the ongoing assembly of the NGC~4839 group.
    We emphasize that this is not driven by the adopted NFW parameterization, since the discrepancy remains with the non-parametric WL mass model.
    The gas mass fraction increases with radius and approaches the cosmic baryon fraction near $R_{200c}$.
    
    \item The one-to-one comparison between the binned WL SNR and X-ray SB shows a noticeable correlation. 
    In addition, the azimuthal WL--X-ray correlation is enhanced along the directions of the reported ICFs, supporting the picture that Coma is accreting matter anisotropically along preferred directions.
   
    \item The NGC~4839 group is clearly detected as a distinct WL subhalo and shows the multiwavelength signatures of an infalling system, including the stripped X-ray tail and radio relic.
    Its gas mass fraction is more consistent with simulated groups near apocenter than with first-infall systems, supporting the scenario that NGC~4839 is currently in its returning phase after a previous core passage.

    \item The cumulative cluster $M/L_r$ remains approximately constant with radius, with an average value of
    $\langle M/L_r \rangle = 250 \pm 66~ M_{\odot} / L_{\odot}$ within $R_{200}$.
    In contrast, the ICFs exhibit substantially higher $M/L_r \gtrsim 1,000~ M_{\odot} / L_{\odot}$ than the cluster average, indicating that these structures are strongly dark matter dominated.
    
\end{itemize}

Overall, our results show that the Coma cluster remains an exceptional nearby laboratory for studying the interplay between dark matter, galaxies, and the ICM during cluster assembly through filaments.
The combination of wide-field WL and multiwavelength data reveals that Coma is not a relaxed monolithic halo, but a dynamically evolving system embedded in a filamentary environment.

\begin{acknowledgements}

KH is supported in part by the Open Universe effort, which is funded by NASA under JPL Contract Task 70-711320, ``Maximizing Science Exploitation of Simulated Cosmological Survey Data Across Surveys" and supported in part by NASA grant 22-ROMAN11-0011, ``Maximizing Cosmological Science with the Roman High Latitude Imaging Survey."
MJJ acknowledges support for the current research from the National Research Foundation (NRF) of Korea under the programs 2022R1A2C1003130 and RS-2023-00219959.
IK was supported by the Simons Foundation via the Simons Investigator Award to A. A. Schekochihin.
W.L. acknowledges support from the National Research Foundation of Korea (NRF) grant funded by the Korea government (MSIT)(RS-2024-00340949).
SC acknowledges this research was supported by Basic Science Research Program through the NRF funded by the Ministry of Education (No. RS-2024-00413036).
HSH acknowledges support from the National Research Foundation of Korea (NRF) funded by the Korea government (MSIT; RS-2026-25482692) and the Global-LAMP Program funded by the Ministry of Education (RS-2023-00301976).
Y.J-T. acknowledges financial support from the State Agency for Research of the Spanish MCIU through Center of Excellence Severo Ochoa award to the Instituto de Astrofísica de Andalucía CEX2021-001131-S funded by MCIN/AEI/10.13039/501100011033, and from the grant PID2022-136598NB-C32 Estallidos and project ref. AST22-00001-Subp-15funded by the EU-NextGenerationEU.
H.C. acknowledges support from the National Research Foundation of Korea (NRF) grants funded by the Korea government (MOE, RS-2022- NR070872; and MSIT, RS-2022-NR070525). 
The Subaru Telescope is operated by the National Astronomical Observatory of Japan. 
We are honored and grateful for the opportunity of observing the Universe from Maunakea, which has the cultural, historical, and natural significance in Hawaii.
This paper makes use of LSST Science Pipelines software developed by the Vera C. Rubin Observatory. 
We thank the Rubin Observatory for making their code available as free software at \url{https://pipelines.lsst.io}.
The calculations with TNG-Cluster in this paper were carried out on the Vera cluster of the Max Planck Institute for Astronomy (MPIA).
This research made use of Photutils, an Astropy package for detection and photometry of astronomical sources \citep{photutils}.

\end{acknowledgements}

\bibliographystyle{aa}
\bibliography{coma}

\end{document}